\documentclass[prl,twocolumn]{revtex4}
\usepackage{dcolumn,amsmath}
\usepackage{bm}
\usepackage{hyperref}   

\newcommand{\eref}[1]{(\ref{#1})}

\newcommand{\tref}[1]{Table~\ref{#1}}

\begin{document}
\title{Towards the electron EDM search. Theoretical study of PbF}
\author{K.I.\ Baklanov}
\author{A.N.\ Petrov$^2$}
\affiliation{Institute of Physics, Saint Petersburg State University, Saint Petersburg, Petrodvoretz 198904, Russia}
\author{A.V.\ Titov}
\author{M.G.\ Kozlov}
\affiliation
{$^2$Petersburg Nuclear Physics Institute, Gatchina,
             Leningrad district 188300, Russia}

\begin{abstract}
 We report \textit{ab initio} relativistic correlation calculations
 of potential curves and spectroscopic constants for four
 lowest-lying electronic states of the
 lead monofluoride.
 We also calculated parameters of the spin-rotational Hamiltonian for the ground
 $^2\Pi_{1/2}$ and the first excited A$^2\Sigma^+_{1/2}$ states including
 \textit{P,T}-odd and \textit{P}-odd terms.
 In particular, we have obtained hyperfine constants of the
 $^{207}$Pb nucleus. For the $^2\Pi_{1/2}$ state
 $A_\perp=-6859.6$~MHz, $A_\|=9726.9$~MHz  and for the
 A$^2\Sigma^+_{1/2}$ $A_\perp=1720.8$~MHz, $A_\|=3073.3$~MHz.
 Our values of the ground state hyperfine constants are in good
 agreement with the previous
 theoretical studies. We discuss and explain seeming disagreement in
 the sign of the constant $A_\perp$ with the recent experimental data.
 The effective electric field on the electron $E_{\rm
 eff}$, which is important for the planned experiment to search for
 the electric dipole moment of the electron,
 is found to be
$3.3 \times 10^{10}{\rm V/cm}$.
\end{abstract}

\maketitle

The interest in theoretical study of $^{207}$PbF is caused by the
planned experiment to search for the simultaneous violation of the
time-reversal invariance (T) and space parity (P) \cite{Kozlov:87,
Dmitriev:92,Shafer-Ray:06,Shafer-Ray:08E}. The latter may be caused
either by the \textit{P,T}-odd electron-nucleus interaction or by
the electric dipole moment of the electron (eEDM, or $d_{\rm e}$),
which can exist also only due to fundamental \textit{P,T}-odd
interactions \cite{Landau:57}. The possible experiments to search
for \textit{P,T}-odd effects in atoms and molecules were discussed
in several books and reviews \cite{Kozlov:95, Khriplovich:97,
Commins:98, Ginges:04, Erler:05, Titov:06amin,Rai08}.

Experiment to search for eEDM with PbF molecule is currently under
preparation by the group of Shafer-Ray \cite{Shafer-Ray:06}. PbF
molecule has one of the strongest effective electric field $E_{\rm
eff}$ on the valence electron, which determines experimentally
measured frequency shifts \cite{Kozlov:87, Dmitriev:92}. This
experiment can test ``new physics'' beyond the Standard Model
including different supersymmetric models. Many such models predict
eEDM in the range between $10^{-28}{-}10^{-29}\ e\cdot {\rm cm}$
($e$ is the electric charge of the electron), within the reach of
the new generation of experiments \cite{ForPatBar03,Rai08}.

Recently \citet{Shafer-Ray:08E} have measured hyperfine structure
parameters for the ground and the first excited states of
$^{207}$PbF and found them to be in disagreement with previous
predictions \cite{Kozlov:87, Dmitriev:92}. Taking into account that
theoretical studies of the lead monofluoride took place decades ago
within rather simple theoretical models we recalculate spectroscopic
constants and parameters of the spin-rotational Hamiltonian, $H_{\rm
sr}$ for PbF molecule. In particular, we focus on the calculations
of the effective electric field on the electron $E_{\rm eff}$ and
the hyperfine structure constants for the lead nucleus. Our new
results are in reasonable agreement with previous calculations
\cite{Kozlov:87, Dmitriev:92}. We discuss the discrepancy between
the theory and experiment \cite{Shafer-Ray:08E} and argue that it is
caused by the incorrect phase factor in the spin-rotational wave
function, rather than in the difference in the sign of the hyperfine
constant $A_\perp$.

 The Hamiltonian $H_{\rm sr}$ can be written as \cite{Dmitriev:92}

\begin{eqnarray}
 \label{Hsr1}
 H_{sr}=B' \textbf J^2+{\Delta} \textbf S'\cdot \textbf
 J+\textbf S' \textbf A \textbf I+ \\
 \label{Hsr2}
 +\mu_{0}\textbf S' \textbf G \textbf B-D\textbf
 {\textit n}\cdot \textbf E+ \\
 \label{Hsr3}
 +W_P\kappa_P\textbf {\textit n} \times\textbf S'\cdot \textbf I
 +(W_{P,T}\kappa_{P,T}+W_d d_e)\textbf S' \cdot \textbf {\textit n},
\end{eqnarray}
where the first line corresponds to the rotational structure with
\textit{$\Omega$}-doubling and the hyperfine interaction of the
effective electron spin ${\bf S'}, S' =\frac{1}{2}$, with the spin
of the nucleus \textbf{I}; $\textit{B}'$ is the rotational constant,
\textbf {J} is the electron-rotational angular momentum, $\Delta$ is
the \textit{$\Omega$}-doubling constant, \textbf A is the hyperfine
structure tensor. The second line describes interaction of the
molecule with the external fields \textbf B and \textbf E, $\bm{n}$
is the unit vector along the molecular axis, $\zeta$, directed from
Pb to F; \textbf G is the G-factor tensor. The last line corresponds
to the interaction with the \textit P-forbidden anapole moment of
the nucleus, the \textit{P,T}-odd weak interactions of the electrons
with the heavy nucleus and the interaction of eEDM with the internal
electric field of the molecule; $\kappa_P$ and $\kappa_{P,T}$ are
the anapole constant and the \textit{P,T}-odd neutral current
constant for Pb nucleus.

Parameters $W_i$ in \eqref{Hsr3} can not be measured experimentally
and have to be found from the molecular electronic structure
calculations. Their accurate knowledge is crucial for extracting
constants $\kappa_P$, $\kappa_{P,T}$, and $d_e$ from the experiment.
Expressions for calculating parameters $W_i$ can be found in
Refs.\cite{Kozlov:87,Kozlov:95,Titov:06amin}. In the literature the
effective electric field seen by the unpaired electron is usually
used instead of the parameter $W_d$. It is defined as: $E_{\rm
eff}=W_d|\Omega|$, where
$\Omega=\langle\Psi^e_\Omega|\bm{S}'\cdot\bm{n}|\Psi^e_\Omega\rangle
= \pm1/2 $ and $\Psi^e_\Omega$ is the electronic wavefunction for
the $^2\Pi_{1/2}$ or $^2\Sigma^+_{1/2}$ states.


 We used the generalized relativistic effective core potential
 (GRECP) \cite{Titov:99}
 to simulate the interaction of valence electrons (4
 electrons of Pb)
 with the explicitly excluded 1\textit s to
 4\textit f electrons of Pb (68 core electrons). In addition the
 5\textit s, 5\textit p and 5\textit d shells of Pb (treated explicitly with
 the used GRECP) were frozen.
  All the electrons of fluorine were treated explicitly.
The resulting configurations were: $6s^26p^2$ for Pb and
$1s^22s^22p^5$ for F. For Pb we used generalized correlation atomic
basis set \cite{Mosyagin:00,Isaev:00} (15\textit s16\textit
p12\textit d9\textit f)/[5\textit s7\textit p4\textit d2\textit f]
and GRECP from the paper \cite{Isaev:00}. For F we used the ANO-I
(14\textit s9\textit p4\textit d3\textit f)/[4\textit s3\textit
p2\textit d1\textit f] atomic basis from the MOLCAS 4.1 library
\cite{MOLCAS}.

By  means of the complete active
 space self-consistent field method the molecular
 orbitals were received. Molecular symmetry was treated by using the
 $C_{2v}$ point group classification scheme.
 The next step was to apply the spin-orbit direct
 configuration interaction (SODCI) approach \cite{Buenker:74, Buenker:99, Alekseyev:04a}
 (modified in \cite{Titov:01}
 to account for spin-orbit interaction in configuration
 selection procedures). The main idea of this method is to use
   selected singly and doubly excited configurations  with respect to some
   multiconfigurational reference states within
 the second order perturbation theory. After that the
 SODCI approach is employed on the space of all the selected configurations.

 In the present work we calculated four lowest states, which belonged to the
 configurations [...]$\sigma ^2\pi_{1/2}$ $(^2\Pi_{1/2})$,
 [...]$\sigma ^2\pi_{3/2}$ $(^2\Pi_{3/2})$, [...]$\sigma ^2\sigma
 '$ (A$^2\Sigma^+_{1/2}$) and [...]$\sigma ^2\sigma ''$
 (B$^2\Sigma^+_{1/2}$). Here $\sigma$ is mainly formed from the
 6\textit s
   basis function
 of Pb, $\pi_{1/2}$ and $\pi_{3/2}$ are mainly
 made from 6$\textit p_x$ and 6$\textit p_y$ functions of Pb. The orbitals
 $\sigma '$ and $\sigma ''$ are formed from the 6$\textit p_z$ and
 7\textit s functions of Pb, respectively.
%
%

One of the main goals of the present paper was to draw potential
curves for the ground and three lowest excited states. Therefore, we
made calculations for several internuclear distances (from 3.8~a.u.\
to 4.2~a.u.) near the ground state equilibrium distance $(R_{\rm
e}=2.0575~\mathrm{\r{A}}\approx 4.0~\rm{a.u.})$.


 Equilibrium internuclear distances, vibrational constants, $T_{\rm e}$ and dipole moments
 for four lowest states of PbF molecule can be found in \tref{un}
 (the most precise data, with 0.00005 treshold, are pointed out
for the $D(R)$).
 The PbF properties could be divided into two groups: valence and
 core ones. The parallel and perpendicular \textbf G-factor
 components, the electronic dipole molecule moment D(R) or $\Omega$--doubling constant
 could be related to
 the valence properties. However, such properties as $E_{\rm
 eff}$, $A_\perp$, $A_\|$, anapole moment constant $W_P$ or \textit T-odd
 interaction constant $W_{P,T}$ could be referred to the core
 ones. Mainly all these properties were calculated for the ground
 and first excited states at the point 4.0 a.u., which is
 close to the equilibrium distance.
 In the core region one should make a four-component
 relativistic calculation.
 However, we have performed calculations on two-component orbitals by means of
 pseudopotential and,
therefore, have to apply a restore procedure for
 four-component orbitals from two-component ones
at the next stage.
 For this purpose the nonvariational one-centre
 restoration method developed in
\cite{Titov:85Dis, Titov:96, Titov:96b, Titov:99, Petrov:02, Petrov:05a}
 is employed.
 One can find the calculated
    parameters of the
 spin-rotational Hamiltonian,
 $H_{sr}$, in \tref{core}
 for the $^2\Pi_{1/2}$ and
 A$^2\Sigma^+_{1/2}$ state, respectively (where (a) and (b) of \cite{Kozlov:87}
 correspond to the minimal and maximal spin-orbital mixing).
 Finally, in
\tref{dipw} the matrix element values of $\Omega$--doubling
constant $\Delta$
divided by
 $2/2MR^2$ ($\Delta/{2\rm B'(R)}$) and electronic dipole moments are
 listed for five investigated points and for two electronic states of the molecule.

\begin{table}[tbh]
\caption{Internuclear distances, vibrational constants, $T_{\rm e}$ and dipole moments for four lowest states of PbF molecule}
\begin{tabular}{|c|c|c|c|c|c|}
\hline
& & $R_{\rm e} (\r{A})$ & $\textit w_{\rm e} (\rm cm^{-1})$ & $T_{\rm e} (\rm cm^{-1})$ & D(R) ({\textit D})\\
\hline
\hline
$^2\Pi_{1/2}$ & \cite{Dmitriev:92} & 2.06 & 489 &  & 4.62\\
& Experiment & 2.0575 & 502.7 & 0 & \\
& This work & 2.08 & 493.6 & 0 & 4.26\\
\hline
\hline
$^2\Pi_{3/2}$ & Experiment & 2.0342 & 528.7 & 8263.5 & \\
& This work & 2.07 & 522.3 & 8258.9 & \\
\hline
\hline
 A$^2\Sigma^+_{1/2}$ & \cite{Dmitriev:92} & 2.18 & 418 &  & 5.5\\
& Experiment & 2.1597 & 394.7 & 22556.5 & \\
& This work & 2.18 & 388.5 & 23380 & 2.51\\
\hline
\hline
B$^2\Sigma^+_{1/2}$ & Experiment & 1.97 & 605.7 & 35644.4 & \\
\hline
& This work & 1.99 & 597.8 & 36583.8 & \\
\hline
\end{tabular}
\label{un}
\end{table}

\begin{table*}[tbh]
\caption{Matrix element values of $\Omega$ - doubling constant
divided to $2B'(R)$ and electronic dipole moments (1\textit D = 0.3934 a.u.) for the ground and the second excited states of PbF molecule}
\begin{tabular}{|c|c|c|c|c|c||c|c|c|c|c|}
\hline
$H_{\rm sr}$ constants &\multicolumn{10}{c|}{$\textless\Psi|\Delta/{2\rm B'}|\Psi\textgreater$}\\
\hline
State & \multicolumn{5}{c||}{$^2\Pi_{1/2}$} & \multicolumn{5}{c|}{A$^2\Sigma^+_{1/2}$}\\
\hline
Internuclear distance & 3.8 & 3.9 & 4.0 & 4.1 & 4.2 & 3.8 & 3.9 & 4.0 & 4.1 & 4.2\\
\hline
\cite{Dmitriev:92} & & & -0.347 & & & & & & & \\
\hline
\cite{Shafer-Ray:08E} & & & -0.303 & & & & & 1.49 & & \\
\hline
This work & -0.412 & -0.342 & -0.369 & -0.391 & -0.411 & 1.49 & 1.42 & 1.45 & 1.47 & 1.49\\
\hline
\hline
 &\multicolumn{10}{c|}{D(R) (\textit D)}\\
\hline
State & \multicolumn{5}{c||}{$^2\Pi_{1/2}$} & \multicolumn{5}{c|}{A$^2\Sigma^+_{1/2}$}\\
\hline
Internuclear distance & 3.8 & 3.9 & 4.0 & 4.1 & 4.2 & 3.8 & 3.9 & 4.0 & 4.1 & 4.2\\
\hline
\cite{Dmitriev:92} & & & 4.62 & & & & & 5.5 & & \\
\hline
This work & 4.27 & 4.264 & 4.28 & 4.266 & 4.255 & 2.32 & 2.42 & 2.55 & 2.67 & 2.79\\
\hline
\end{tabular}
\label{dipw}
\end{table*}

 According to the data pointed out in \tref{un}, $T_{\rm e}$ values
 are in good agreement with the experimental data.  The obtained equilibrium
 distance, $R_{\rm e}$, have slightly higher value than in the experiment,
 whereas the presently obtained vibrational constants are in much better
 agreement with the experiment
 comparing to those from \cite{Dmitriev:92}. Note that the dipole moment $D(R)$
 value for the ground state are nearly the same to those obtained in
\cite{Dmitriev:92},
 whereas they are decreased approximately two times for the
 excited state. All data
in \tref{dipw} obtained for D(R) and \textit{$\Omega$}-doubling constant
have been calculated with the 0.0005 threshold. But we should note that we
 have computed these parameters for point 4.0 a.u. more accurately with
 0.00005 threshold. For \textit{$\Omega$}-doubling these values are -0.361
 and 1.46 for the $^2\Pi_{1/2}$ and A$^2\Sigma^+_{1/2}$ states, respectively.

 Let us thoroughly examine the data obtained in
\tref{core}.
 First of
 all we would like to discuss the results for the ground state. The first
 point to note is that the $A_\|$ value is in the
 fault bounds, meantime the $G_\|$ value is increased in two times according to
 \cite{Dmitriev:92}.
 The effective field constant $W_d$ obtained is the best datum to-date,
 the anapole moment constant $W_P$ and \textit T-odd
 interaction constant $W_{P,T}$ increased their values nearly by
 $50\%$. The most interesting result
concerns the
perpendicular hyperfine structure constant $A_\perp$.
The sign of the obtained value exactly agree with those in previous
works \cite{Kozlov:87, Dmitriev:92}.
Since
the constants of the
$H_{sr}$ such as
$A_\perp$, $G_\perp$, \textit{$\Omega$}-doubling
are determined by the off-diagonal matrix element,
it is important to fix the phase factors for the wave
functions to get the consistent signs of those matrix elements.
In our calculations the phase factor is the same as in eq.~(14) of \cite{Kozlov:87}.

 In the experimental work
\cite{Shafer-Ray:08E} of the Shafer-Ray group
 the obtained
 $A_\perp$ values are nearly the same by absolute value but have the
 opposite sign for both states. However, it was noted in \cite{Shafer-Ray:08E}
that the energy level
 structure is determined by the expression, in which $A_\perp$ is
 multiplied by $p$, where $p = \pm 1$ is the parity of the
 electronic-rotational state. Therefore, it was suggested that
 the parity rather than sign of $A_\perp$ has been
incorrectly assigned in previous works. However, we believe that
the changes
    in parity or sign of the perpendicular component
are not necessary.
Some corrections for the energy level expression are required instead.
As we found, $(-1)^{J+1/2}$ should be replaced by $(-1)^{J-1/2}$ in
eqs. (9) and (23) of \cite{Kozlov:87}.
This leads to change of sign in front of
$pA_\perp$ in eq.~(24) of \cite{Kozlov:87} for the energy level structure.

Let us consider the symmetry properties of wavefunction
of a diatomic molecule at
   the space inversion operation
in more details.
Let
%
$x,y,z$ be a
  laboratory axes,
%
and the $\xi,\eta,\zeta$ body frame axes
in which the molecule is fixed. The $\alpha$ and $\beta$ ($\gamma=0$)
Euler angles are defined in such a way that the $\zeta$ axis coincides with the
molecule axis.
The
vibrational wavefunction is not important here,
and
electronic-rotational wavefunction within the adiabatic approximation in the Hund case
$c$ is written as
\begin{equation}
\Psi^{e}_{\Omega}\left(\ldots,\xi_a,\eta_a,\zeta_a,\tilde{\sigma}_a,\ldots,R
\right)\theta^{J}_{M_J,\Omega}(\alpha,\beta),
\label{adiab}
\end{equation}
 where $M_J$ is projection of the total angular momentum, $J$, on the $z$ axis,
subscript $a$ enumerate electrons,
$\tilde{\sigma}_a=\pm 1$ is the spin variable which corresponds to
the projection of the spin of the $a$-th\- electron on the $\zeta$
axis, $R$ is the internuclear distance.
$\theta^{J}_{M_J,\Omega}(\alpha,\beta)=\sqrt{\frac{2J+1}{4\pi}}D^{J}_{M_J,\Omega}(\alpha,\beta,\gamma=0)$
is rotational wavefunction.
In the laboratory system of coordinates the inversion
operation,$\hat{P}$, is
\begin{eqnarray*}
\hat{P}\tilde{\Psi}^{e}_{\Omega}\left(\ldots,x_a,y_a,z_a,\sigma_a\ldots,R,\alpha,\beta
\right)\theta^{J}_{M_J,\Omega}(\alpha,\beta) = \\
\nonumber
\tilde{\Psi}^{e}_{\Omega}\left(\ldots,-x_a,-y_a,-z_a,\sigma_a\ldots,R,\alpha+\pi,\pi-\beta
\right)\\
\times\theta^{J}_{M_J,\Omega}(\alpha+\pi,\pi-\beta),
\end{eqnarray*}
where $\Psi^{e}_{\Omega} = \hat{U}\tilde{\Psi}^{e}_{\Omega}$,
$\hat{U}=exp\left(i\beta J^e_2 \right)\cdot exp\left(i\alpha J^e_3
\right)$ is the unitary operator. Then one can obtain that the inversion,
$\hat{\tilde{P}}=\hat{U}\hat{P}\hat{U}^+$, in the
body frame
system of coordinates is
\begin{eqnarray}
\nonumber
\hat{\tilde{P}}\Psi^{e}_{\Omega}\left(\ldots,\xi_a,\eta_a,\zeta_a,\tilde{\sigma}_a,\ldots,R
\right)\theta^{J}_{M_J,\Omega}(\alpha,\beta) = \\
\nonumber
i^N\Psi^{e}_{\Omega}\left(\ldots,-\xi_a,\eta_a,\zeta_a,-\tilde{\sigma}_a,\ldots,R
\right) \\
\times\theta^{J}_{M_J,\Omega}(\alpha+\pi,\pi-\beta),
\end{eqnarray}
 where $i=\sqrt{-1}$, $N$ is the number of electrons.
Note that the above transformation of the electronic wavefunction
(corresponding to the reflection in the  $\eta-\zeta$ plane) gives the electronic
wavefunction with the opposite sign of $\Omega$, which has the same adiabatic
potential. In the case of $\Omega=0$ electronic wavefunction whether not changed
or is multiplied by the $-1$ factor.
The rotational wavefunctions are also transformed into each other with
changing sign of $\Omega$ \cite{LL77}:
\begin{equation}
\theta^{J}_{M_J,\Omega}(\alpha+\pi,\pi-\beta)=
\left(-1 \right)^{J+2\Omega}\theta^{J}_{M_J,-\Omega}(\alpha,\beta)
\label{cond1}
\end{equation}
The total electronic-rotational wavefunction (\ref{adiab}), except the
case $\Omega=0$, has no definite parity. In the case $\Omega\ne0$ two
wavefunctions (\ref{adiab}) with different signs of $\Omega$
are transformed into each other, and we can form from them positive ($p=1$)
and negative ($p=-1$) wavefunctions. Let us construct the
electronic-rotational wavefunction with definite parity for the
case $\Omega=\pm1/2$ of our interest, and, in particular, for
the $^2\Pi_{1/2}$ and $^2\Sigma^+$ states. Since the closed electronic shell is
invariant under reflection, consider transformation only of the
valence spinor
$\varphi^e_{\omega}(\xi_a,\eta_a,\zeta_a,\tilde{\sigma})$, where $\omega=\pm1/2$ is
projection of the total angular momentum, $j$, of the valence electron on the $\zeta$
 axis. One can verify that for phase factors accepted by eq. (14) of \cite{Kozlov:87}
\begin{equation}
i\varphi^e_{1/2}(-\xi_a,\eta_a,\zeta_a,-\tilde{\sigma})=
i\varphi^e_{-1/2}(\xi_a,\eta_a,\zeta_a,\tilde{\sigma})
\label{cond2}
\end{equation}
and hence, accounting for \eref{cond1} and \eref{cond2}, the
electronic-rotational wavefunction $\left|J,M_J,p\right>$ with parity $p$
   becomes
\begin{eqnarray}
\nonumber
\left|J,M_J,p\right> = \\
\Psi^{e}_{1/2}\theta^{J}_{M_J,1/2}+
(-1)^{J-1/2}p\Psi^{e}_{-1/2}\theta^{J}_{M_J,-1/2}
\label{par}
\end{eqnarray}
Eq.~(23) of \cite{Kozlov:87} is reduced to \eref{par} by
replacing $(-1)^{J+1/2}$
on $(-1)^{J-1/2}$.

 The parallel hyperfine structure components is in agreement with
 \cite{Shafer-Ray:08E}.
The other data for the first excited A$^2\Sigma^+_{1/2}$ state,
listed in the \tref{core}, were obtained for the first time.

   Finally, we would like to note that we have performed the most accurate to
   date electronic structure calculation of the lowest states of PbF and, as a result,
   have obtained the most reliable data for the spin-rotational Hamiltonian constants.
In this paper we have
   proposed solution of
the problem of the sign mismatching in the
perpendicular
hyperfine structure constant.  The new computed
data for the spectroscopic constants better conform to the experimental results
at present.  The calculated effective electric field on the electron is
increased in the present study nearly by $10\%$ compared to that in
\cite{Dmitriev:92}.

\begin{table*}[tbh]
\caption{Parameters of the spin-rotational Hamiltonian $H_{\rm sr}$ for $^2\Pi_{1/2}$ and A$^2\Sigma^+_{1/2}$ states of PbF molecule}
\begin{tabular}{|c|c|c|c|c|c|c|c|c|}
\hline
State & Reference & $A_\perp$ (MHz) & $A_\parallel$ (MHz) & $G_\perp$ & $G_\parallel$ & $W_d(10^{25}){\rm Hz}/e~{\rm cm}$ & $W_{P,T}$(KHz) & $W_P$(KHz) \\
\hline
\hline
$^2\Pi_{1/2}$ & \cite{Dmitriev:92} & -8990 & 10990 & -0.326 & 0.040 & 1.4 & 55 & -0.72\\
& \cite{Shafer-Ray:08E} & $7200\pm 150$ & $10300\pm 800$ & & & & & \\
& \cite{Meyer:08} & & & & & 1.5 & & \\
& \cite{Kozlov:87} (a) & -7460 & 8690 & -0.269 & 0.034 & 1.0 & 51 & -0.65\\
& \cite{Kozlov:87} (b) & -8240 & 9550 & -0.438 & 0.114 & 1.8 & 99 & -1.25\\
& This work & -6859.6 & 9726.9 & -0.319 & 0.082 & 1.6 & 74.95 & -0.99\\
\hline
\hline
A$^2\Sigma^+_{1/2}$ & \cite{Shafer-Ray:08E} & $-1200\pm 300$ & $3000\pm 2500$ & & & & & \\
& This work & 1720.8 & 3073.3 & 2.417 & 1.92 & 2.5 & 123.7 & -1.588\\
\hline
\end{tabular}
\label{core}
\end{table*}

K.B. and A.P.\ are grateful to the Ministry of Education and Science of Russian
Federation (Program for Development of Scientific Potential of High School) for
the Grant No.~2.1.1/1136.  A. T.\ and A.P.\ were supported by the RFBR grant
09--03--01034


\end{document}